\documentclass[11pt,twoside,onecolumn]{article}
\usepackage[]{latexsym}
\usepackage{epsfig}
\usepackage{amsmath,amssymb}
\newcommand{\be}{\begin{eqnarray}}
\newcommand{\ee}{\end{eqnarray}}

\title{\bf Tolman IV solution in the\\ Randall-Sundrum Braneworld}
\author{J Ovalle,$^{a}$\thanks{Corresponding author: jovalle@usb.ve}
$\,$ 
F Linares,$^{b}$\thanks{fran2012@fisica.ugto.mx}
\\
\null
\\
$^a${\em Departamento de F\'{\i}sica, Universidad Sim\'on Bol\'ivar}
\\
{\em Apartado 89000, Caracas 1080A, Venezuela}
\\
\\
$^b${\em Departamento de F\'{\i}sica, Universidad de Guanajuato, Le\'on, M\'exico}
}
\begin{document}
\maketitle
\begin{abstract}
In the context of the Randall-Sundrum braneworld, the minimal geometric deformation approach (MGD) 
is used to generate an exact analytic interior solution to four-dimensional effective Einstein's field equations
for a spherically symmetric compact distribution. This solution represents the braneworld version 
of the well known Tolman IV solution in General Relativity. By using this analytic solution, an exhaustive analysis of the 
braneworld effects on realistic stellar interiors is developed, finding strong evidences in favor of the hypothesis that compactness 
is reduced due to bulk effects on stellar configurations.

\end{abstract}
\iffalse
\begin{abstract}
In the context of the Randall-Sundrum braneworld, the minimal geometric deformation approach (MGD) 
is used to generate a physically acceptable exact interior solution to Einstein's field equations
for a spherically symmetric compact distribution. This exact solution represents the braneworld version 
of the well known Tolman IV solution in General Relativity. By ussing this exact solution, 
braneworld effects on realistic stellar interiors are investigated.
\end{abstract}
\fi
%
%\keywords{Brane-world, Stars, Black holes}
%\pacs{04.50.+h, 04.70.-s, 04.70.Dy}
%
%
%
%
%
%
\section{Introduction}
\setcounter{equation}{0}
\label{intro}
In recent years there has been great interest in finding alternative theories to General Relativity (GR) \cite{beyond, capo1}, mainly due to the inability of the 
latter to explain satisfactorily some fundamental issues associated with the gravitational interaction, such as the dark matter problem, the dark energy problem, as well as the impossibility to reconcile GR with the Standard Model of particle physics. Extra-dimensional theories,
which are mostly inspired by String/M-theory, are among the theories that lead to gravity beyond GR. One of 
these extra-dimensional theories is the Braneworld (BW) proposed by Randall and Sundrum (RS) \cite{RS} which has been largely studied and which explains 
one of the fundamental problems of Physics, i.e. the hierarchy problem (see also the ADD model \cite{ADD} and \cite{AADD}).  Because of this, its study and impact on GR is fully justified and is of great importance 
\cite{maartRev2004}.

Even though we have a covariant approach that is useful to study many fundamental aspects of the theory of RS BW
\cite{SMS}, we are still far from fully understanding its 
impact on gravity, mainly due to the lack of the complete five-dimensional solution (bulk plus brane), which could helps to explain certain key issues 
that remain unresolved, such as the existence of black holes in RS BW \cite{FW11}-\cite{kanti2013} and the bulk effects on stellar configurations \cite{germ}. Since the complete five-dimensional solution remains unknown so far, finding exact solutions to four-dimensional effective Einstein field equations in the 
brane is a convenient way to clarify some aspects of the five-dimensional geometry, essentially because we could use Campbell-Magaard 
theorems \cite{campbell, sss} to extend the brane solution through the bulk, 
locally at least. However, GR during its almost century of history, has taught us that to find a physically acceptable exact solution of Einstein's field equations 
is an extremely difficult task \cite{Stephani}. This mainly due to the complexity of the field equations. If we deal with internal stellar 
solutions \cite{lake03herrera08}, the task is much more complicated, and in fact, just a few internal solutions are known \cite{lake98}. 
On the other hand, in the context of BW, two important features, completely new and different from GR, greatly complicate 
the searching for solutions to 4-dimensional Einstein's field equations in the astrophysical scenario: 
1) The system remains indefinity due to nonlocal corrections from the five-dimensional bulk. 2) The presence of nonlinear 
terms in material fields due to high energy corrections \cite{maartRev2004, SMS}. Because the latter, 
to find exact and physically acceptable stellar interior solutions to effective 
4-dimensional Einstein field equations seems an impossible task to carry out. However, these two problems can be solved {\it simultaneously on the brane} 
when a GR solution is considered by using the minimal geometric deformation principle (MGD) \cite{jovalle2009}. Indeed, by using this approach, 
an exact and physically acceptable solution on the brane was found in Ref. \cite{jovalle207}. The MGD  has allowed, among other things, to generate physically acceptable interior 
solutions for stellar systems \cite{jovalleBWstars}, to solve the tidally charged exterior solution found in Ref \cite{dadhich} in terms of the 
ADM mass and to study (micro) black hole solutions \cite{covalle1, covalle2}, as well as to help to elucidate the role of exterior Weyl stresses
from bulk gravitons on compact stellar distributions \cite{olps2013} and the behaviour of black string models with variable brane 
tension \cite{cor2013}.

In this paper, an analytical solution to Einstein field equations for a non uniform stellar structure is 
found on the brane, and used to elucidate the effects of bulk gravitons on compact stellar structures. 
The MGD approach will be used to modify the perfect fluid solution represented by a well 
known general relativistic solution, namely, the Tolman IV solution \cite{tolman}, generating thus its braneworld version in an exact analytical form. 
The reason to investigate the Tolman IV solution in the Braneworld context by using the MGD approach is quite obvious: among hundreds of known exact solutions 
in GR, the Tolman IV solution is one of few with physical meaning \cite{phymeaning}, and this physical relevance is naturally inherited by its braneworld version. 

This paper is organized as follows. 
In Section {\bf 2} the Einstein field equations in the brane for a spherically symmetric and static distribution of density $\rho$ and pressure $p$ 
is reminded. In Section {\bf 3} the MGD approach is discussed, as well as the general matching conditions between an interior deformed 
metric and the exterior one associated to a Weyl fluid with dark pressure  ${\cal P}^+$ and dark 
radiation ${\cal U}^+$. In Section {\bf 4} an analytical stellar interior solution to the effective 4-dimensional Einstein's fields 
equations is generated by using the well known Tolman IV GR solution through the MGD approach. In Section {\bf 5} the far-field correction to the Newtonian potencial in the BW is used to construct an exterior geometry associated with this potential. In this approximation, the bulk effects on stellar configurations is elucidated. In the last section 
the conclusions are presented.

\section{General framework}

In the context of the braneworld, the five-dimensional gravity produces a modification on Einstein's field equations in our (3+1)-dimensional observable universe, 
the so-called brane, which effectively can be written as follow
\begin{equation}
\label{einst}
G_{\mu\nu}=
-k^2\,T_{\mu\nu}^{T}-\Lambda\, g_{\mu\nu}
\ ,
\end{equation}
where $k^2=8\,\pi\,G_{\rm N}$ and $\Lambda$ is the cosmological constant
on the brane. These modifications can be seen through the effective energy-momentum tensor $T_{\mu\nu}^{T}$, 
which has new terms carrying five-dimensional consequences onto the brane:
\begin{equation}\label{tot}
T_{\mu\nu}\rightarrow T_{\mu\nu}^{\;\;T}
=T_{\mu\nu}+\frac{6}{\sigma}S_{\mu\nu}+\frac{1}{8\pi}{\cal
E}_{\mu\nu}+\frac{4}{\sigma}{\cal F}_{\mu\nu},
\end{equation}
where $\sigma$ is the brane tension, with  $S_{\mu\nu}$ and
$\cal{E}_{\mu\nu}$ the high-energy and  non-local (from the point of view of a brane observer) corrections respectively, and ${\cal F}_{\mu\nu}$ a term which depends on all stresses in the bulk but the cosmological constant. In this paper, only the cosmological constant will be considered in the bulk, hence ${\cal F}_{\mu\nu}=0$, which implies there will be no exchange of energy between the bulk and the brane, and therefore $\nabla^\nu\,T_{\mu\nu}=0$. 
The high-energy $S_{\mu\nu}$
and Kaluza-Klein $\cal{E}_{\mu\nu}$ corrections are given by
\be
\label{s}
S_{\mu\nu}=
\frac{T\,T_{\mu\nu}}{12}
-\frac{T_{\mu\alpha}\,T^\alpha_{\ \nu}}{4}
+\frac{g_{\mu\nu}}{24}
\left[3\,T_{\alpha\beta}\,T^{\alpha\beta}-T^2\right]
\ee
where $T=T_\alpha^{\ \alpha}$, and
\be
\label{e}
k^2\,{\cal E}_{\mu\nu}
=
\frac{6}{\sigma}\left[{\cal U}\left(u_\mu\,u_\nu+\frac{1}{3}\,h_{\mu\nu}\right)
+{\cal P}_{\mu\nu}+{\cal Q}_{(\mu}\,u_{\nu)}\right]
\ ,
&&
\ee
with ${\cal U}$, ${\cal P}_{\mu\nu}$ and
${\cal Q}_\mu $ the bulk Weyl scalar, the anisotropic stress and energy flux, respectively, 
and $u^\mu$ the cuadrivelocity with $h_{\mu\nu}=g_{\mu\nu}-u_{\mu}u_{\nu}$ the projection tensor.
In this paper we will consider spherically symmetric static distributions,
hence $Q_\mu =0$ and
\begin{equation}
{\cal P}_{\mu\nu}
={\cal P}\left(r_\mu\, r_\nu+\frac{1}{3}\,h_{\mu\nu}\right)
\  ,
\end{equation}
where $r_\mu$ is a unit radial vector. Furthermore, the line element will be given by Schwarzschild-like coordinates
\begin{equation}
\label{metric}ds^2=e^{\nu(r)} dt^2-e^{\lambda(r)} dr^2-r^2\left( d\theta
^2+\sin {}^2\theta d\phi ^2\right)\, ,
\end{equation}
where $\nu=\nu(r)$ and $\lambda=\lambda(r)$ are functions of
the areal radius $r$, which ranges from $r=0$ (the star's centre)
to $r=R$ (the star's surface). In this paper we will be focused in BW consequences on perfect fluids, hence
the energy-momentum tensor $T_{\mu\nu}$ in Eq. (\ref{tot}) corresponds
to a perfect fluid, given by
\be
\label{perfect}
T_{\mu\nu}
=
(\rho+p)\,u_\mu\,u_\nu-p\,g_{\mu\nu}
\ ,
\ee
where $u^\mu=e^{-\nu/2}\,\delta_0^\mu$ is the fluid four-velocity field in the
reference frame where the metric takes the form in Eq.~(\ref{metric}) (for early works on astrophysics in the braneworld context, 
see for instance Refs. \cite{CFMsolution}-\cite{Gregory2006}). 
\par
The metric (\ref{metric}) must satisfy the effective 4-D Einstein field
equations (\ref{einst}), which, for $\Lambda=0$, explicitly read (For details, see Ref. \cite{covalle2})
%\begin{widetext}
\begin{eqnarray}
\label{ec1}
&&
k^2
\left[ \rho
+\strut\displaystyle\frac{1}{\sigma}\left(\frac{\rho^2}{2}+\frac{6}{k^4}\,\cal{U}\right)
\right]
=
\strut\displaystyle\frac 1{r^2}
-e^{-\lambda }\left( \frac1{r^2}-\frac{\lambda'}r\right)
\\
\nonumber
\\
&&
\label{ec2}
k^2
\strut\displaystyle
\left[p+\frac{1}{\sigma}\left(\frac{\rho^2}{2}+\rho\, p
+\frac{2}{k^4}\,\cal{U}\right)
+\frac{4}{k^4}\frac{\cal{P}}{\sigma}\right]
=
-\frac 1{r^2}+e^{-\lambda }\left( \frac 1{r^2}+\frac{\nu'}r\right)
\\
\nonumber
\\
&&
\label{ec3}
k^2
\strut\displaystyle\left[p
+\frac{1}{\sigma}\left(\frac{\rho^2}{2}+\rho\, p
+\frac{2}{k^4}\cal{U}\right)
-\frac{2}{k^4}\frac{\cal{P}}{\sigma}\right]
=
\frac 14e^{-\lambda }\left[ 2\,\nu''+\nu'^2-\lambda'\,\nu'
+2\,\frac{\nu'-\lambda'}r\right] 
\ .
\nonumber
\\
\end{eqnarray}
%\end{widetext}
Moreover,
\be
\label{con1}
p'=-\strut\displaystyle\frac{\nu'}{2}(\rho+p)
\ ,
\ee
where $f'\equiv \partial_r f$.
We then note that four-dimensional GR equations are formally
recovered for $\sigma^{-1}\to 0$, and the conservation equation~(\ref{con1})
then becomes a linear combination of Eqs.~(\ref{ec1})-(\ref{ec3}).
\par
The Israel-Darmois matching conditions~\cite{israel} at the stellar surface
$\Sigma$ of radius $r=R$ give
\be
\label{matching1}
\left[G_{\mu\nu}\,r^\nu\right]_{\Sigma}=0
\ ,
\ee
where $[f]_{\Sigma}\equiv f(r\to R^+)-f(r\to R^-)$.
Using Eq.~\eqref{matching1} and the general field equations~\eqref{einst},
we find
\be
\label{matching2}
\left[T^{T}_{\mu\nu}\,r^\nu\right]_{\Sigma}=0
\ ,
\ee
which in our case leads to
\be
\label{matching3}
\left[
p+\frac{1}{\sigma}\left(\frac{\rho^2}{2}+\rho\, p
+\frac{2}{k^4}\,\cal{U}\right)+\frac{4}{k^4}\,\frac{\cal{P}}{\sigma}
\right]_{\Sigma}=0
\ .
\ee
Since we assume the distribution is surrounded by a Weyl fluid ${\cal U}^+, {\cal P}^+$,
$p=\rho=0$ for $r>R$, this matching condition
takes the final form
%\begin{widetext}
\be
\label{matchingf}
p_R+\frac{1}{\sigma}\left(\frac{\rho_R^2}{2}+\rho_R\, p_R
+\frac{2}{k^4}\,{\cal U}_R^-\right)
+\frac{4}{k^4}\frac{{\cal P}_R^-}{\sigma}
=
\frac{2}{k^4}\frac{{\cal U}_R^+}{\sigma}+\frac{4}{k^4}\frac{{\cal P}_R^+}{\sigma}
\ ,
\ee
%\end{widetext}
where $f_R^\pm\equiv f(r\to R^\pm)$, with $p_R\equiv p_R^-$
and $\rho_R\equiv \rho_R^-$.
\par
Eq.~\eqref{matchingf} gives a general matching condition
for any static spherical BW star~\cite{germ,gergely2006}, i.e., the second fundamental form. 
In the limit $\sigma^{-1}\rightarrow 0$, we obtain the well-known GR
matching condition $p_R =0$ at the star surface.
In the particular case of the Schwarzschild exterior,
${\cal U}^+={\cal P}^+ =0$, the matching condition~\eqref{matchingf}
becomes
\be
\label{matchingfS}
p_R+\frac{1}{\sigma}\left(\frac{\rho_R^2}{2}+\rho_R\, p_R
+\frac{2}{k^4}\,{\cal U}_R^-\right)
+\frac{4}{k^4}\frac{{\cal P}_R^-}{\sigma} = 0
\ .
\ee
This clearly shows that, because of the presence of
${\cal U}_R^-$ and ${\cal P}_R^-$, the matching conditions
do not have a unique solution in the BW.
\par

\section{Star interior and geometric deformation}
Two important aspects regarding the system of Eqs. (\ref{ec1})-(\ref{con1}) are worth being highlighted. First of all, 
it represents an indefinite system of
equations in the brane, an open problem for which the solution requires  more information of the bulk geometry and a
better understanding of how our four-dimensional spacetime is embedded in the bulk \cite{FW11}-\cite{LBH13}, \cite{cmazza}-\cite{darocha2012}. Secondly, to find exact and physically acceptable analytic functions $(\rho, p, \lambda, \nu, {\cal U}, {\cal P})$ being a solution of the system 
(\ref{ec1})-(\ref{con1}), seems an impossible task.  Even though the second point is quite obvious, we will see that it is possible 
to build an exact and physically acceptable solution by using  the MGD approach \cite{jovalle2009}. In order to accomplish this, the first 
step is to rewrite field equations~\eqref{ec1}-\eqref{ec3} as follow
\be
\label{usual}
&&
e^{-\lambda}
=
1-\frac{k^2}{r}\int_0^r
x^2
\left[
\rho+\frac{1}{\sigma}\left(\frac{\rho^2}{2}+\frac{6}{k^4}\,\cal{U}\right)
\right]
dx\ ,
\\
\label{pp}
\nonumber
\\
&&
\frac{1}{k^2}\,\frac{{\cal P}}{\sigma}
=
\frac{1}{6}\left(G_{\ 1}^{1}-G_{\ 2}^2\right)\ ,
\\
\label{uu}
\nonumber
\\
&&
\frac{6}{k^4}\,\frac{{\cal U}}{\sigma}
=
-\frac{3}{\sigma}\left(\frac{\rho^2}{2}+\rho\,p\right)
+\frac{1}{k^2}\left(2\,G_{\ 2}^2+G_{\ 1}^1\right)-3\,p
\ ,
\ee
with
\be
\label{g11}
G_{\ 1}^1
=
-\frac 1{r^2}+e^{-\lambda }\left( \frac 1{r^2}+\frac{\nu'}r\right)\ ,
\ee
and
\be
\label{g22}
G_{\ 2}^2
=
\frac 14\,e^{-\lambda }\left( 2\,\nu''+\nu'^2-\lambda'\,\nu'+2 \frac{\nu'-\lambda'}r
\right)
\ .
\ee 
\par
Now, by using Eq.~\eqref{uu} in Eq.~\eqref{usual} an integro-differential equation for the function
$\lambda=\lambda(r)$ is found, something completely different from the GR case,
and a direct consequence of the non-locality of the BW equations.
The only general solution known for this equation is given by~\cite{jovalle2009}
%\begin{widetext}
\be
\label{edlrwss}
e^{-\lambda}
&\!\!=\!\!&\underbrace{
{1-\frac{k^2}{r}\int_0^r
x^2\,\rho\,dx}}_{\rm GR-solution}
+\underbrace{e^{-I}\int_0^r\frac{e^I}{\frac{\nu'}{2}+\frac{2}{x}}
\left[H(p,\rho,\nu)+\frac{k^2}{\sigma}\left(\rho^2+3\,\rho \,p\right)\right]
dx+\beta(\sigma)\,e^{-I},}_{\rm Geometric\ deformation}
\nonumber
\\
&\!\!\equiv\!\!&
\mu(r)+f(r)
\ ,
\ee
where
\be
\label{finalsol}
H(p,\rho,\nu)
\equiv
3\,k^2\,p
-\left[\mu'\left(\frac{\nu'}{2}+\frac{1}{r}\right)
+\mu\left(\nu''+\frac{\nu'^2}{2}+\frac{2\nu'}{r}+\frac{1}{r^2}\right)
-\frac{1}{r^2}\right]
\ ,
\ee
%\end{widetext}
and
\be
\label{I}
I
\equiv
\int\frac{\left(\nu''+\frac{{\nu'}^2}{2}+\frac{2\nu'}{r}+\frac{2}{r^2}\right)}
{\left(\frac{\nu'}{2}+\frac{2}{r}\right)}\,dr
\ ,
\ee
with $\beta(\sigma)$ a function of the brane tension $\sigma$ which must be zero in the GR limit. In the case of interior solutions, 
the condition $\beta(\sigma)=0$ has to be imposed to avoid singular solutions at the center $r=0$.
Note that the function 
\be
\label{standardGR}
\mu(r)
\equiv
1-\frac{k^2}{r}\int_0^r x^2\,\rho\, dx
=1-\frac{2\,m(r)}{r}
\ee
contains the usual GR mass function $m$,
whereas the function $H(p,\rho,\nu)$ encodes anisotropic effects due to bulk gravity consequences on $p$,
$\rho$ and $\nu$ .
\par
A crucial observation is now that, when a given (spherically symmetric) perfect fluid solution in GR is considered
as a candidate solution for the BW system of Eqs.~\eqref{ec1}-\eqref{con1}
[or, equivalently, Eq.~\eqref{con1} along with Eqs.~\eqref{usual}-\eqref{uu}],
one obtains
\be
H(p,\rho,\nu)=0
\ ,
\label{H=0}
\ee
therefore every (spherically symmetric) perfect fluid solution in GR
will produce a {\it minimal} deformation on the radial metric component (\ref{edlrwss}), given by
\be
\label{fsolutionmin}
f^{*}(r)
=
\frac{2\,k^2}{\sigma}\,
e^{-I(r)}\int_0^r
\frac{x\,e^{I(x)}}{x\,\nu'+4}\left(\rho^2+3\,\rho\, p\right)
dx
\ .
\ee
The expression given by Eq.~(\ref{fsolutionmin}) represents a minimal
deformation in the sense that all sources of the deformation in (\ref{edlrwss}) have been removed,
except for those produced by the density and pressure, which will always
be present in a realistic stellar  distribution~\footnote{There is a MGD solution
in the case of a dust cloud, with $p=0$, but we will not consider it in the present work.}.
It is worth emphasising that the geometric deformation $f(r)$ shown in
Eq.~\eqref{edlrwss} indeed ``distorts'' the GR solution given in Eq.~\eqref{standardGR}.
The function $f^{*}(r)$ shown in Eq.~\eqref{fsolutionmin} will therefore produce,
from the GR point of view, a ``minimal distortion'' for any GR solution
one wishes to consider, being this distortion $f^{*}(r)$ the source of the anisotropy induced in the brane, 
whose explicit form may be found through Eq. (\ref{pp}), leading to
\begin{eqnarray}
\label{ppf3}
\frac{48\pi}{k^4}\frac{{\cal P}}{\sigma} =
\bigg(\frac{1}{r^2}+\frac{\nu'}{r}\bigg)f^{*}
-\frac{1}{4}\bigg(2\nu''+{\nu'}^2+2\frac{\nu'}{r}\bigg)f^{*}-\frac{1}{4}\bigg(\nu'+\frac{2}{r}\bigg){(f^{*})}'.
\nonumber \\
\end{eqnarray}
It is clear that this minimal deformation will produce a minimal anisotropy onto the brane. 

In this approach, the interior stellar geometry is generically described by the MGD metric, which explicitly read
\begin{equation}
\label{mgdmetric} 
ds^2=e^{\nu(r)} dt^2-\frac{dr^2}{\left(1-\frac{2\,m(r)}{r}+f^*(r)\right)}-r^2\left( d\theta
^2+\sin {}^2\theta d\phi ^2\right)\, .
\end{equation}
As it is shown by Eq. (\ref{fsolutionmin}), the geometric deformation $f^{*}(r)$ in Eq. (\ref{mgdmetric}) satisfies $f^{*}(r)\geqslant\,0$, hence it always reduces the effective interior mass, 
as it is seen further below in Eqs. (\ref{reglambda}) and (\ref{massfunction}).

\subsection{Matching conditions: interior MGD metric and exterior Weyl fluid.}

The MGD metric in (\ref{mgdmetric}), characterizing the interior stellar $r<R$, must be matched with a exterior 
solution associated to the Weyl fluid ${\cal U}^+, {\cal P}^+$,
$p=\rho=0$ for $r>R$, which can be written generically as
\begin{equation}
\label{genericext}ds^2=e^{\nu^+(r)} dt^2-e^{\lambda^+(r)} dr^2-r^2\left( d\theta
^2+\sin {}^2\theta d\phi ^2\right)\, ,
\end{equation}
therefore the continuity of the first fundamental form at the stellar surface $r=R$
\begin{equation}
\label{match1} 
\left[ds^2\right]_{\Sigma}=0
\end{equation}
leads to
\begin{eqnarray}
 \label{ffgeneric1}
e^{\nu^-(R)}&=&e^{\nu^+(R)}\ ,
\\
 \label{ffgeneric2}
1-\frac{2\,M}{R}+f^*_R&=&e^{-\lambda^+(R)}\ ,
\end{eqnarray}
whereas the second fundamental form (\ref{matchingf}) leads to
\begin{equation}
 \label{sfgeneric}
p_R+\frac{f^*_R}{8\pi}\left(\frac{\nu'_R}{R}+\frac{1}{R^2}\right)= \frac{2}{k^4}\frac{{\cal
U}_R^+}{\sigma}+\frac{4}{k^4}\frac{{\cal P}_R^+}{\sigma}\ .
\end{equation}
The expressions given by Eqs. (\ref{ffgeneric1})-(\ref{sfgeneric}) are the necessary and sufficient conditions for the matching of the MGD metric to a spherically symmetric ``vaccum'' filled by a BW Weyl fluid.

\section{An interior solution.} 

As already was mentioned, the system of Eqs. (\ref{ec1})-(\ref{con1}) [or, equivalently, Eq.~\eqref{con1} along with Eqs.~\eqref{usual}-\eqref{uu}], represents an indefinite system of
equations in the brane, an open problem whose answer requires the complete five-dimensional solution. Given that there is no such a solution, the first obvious question is to ask what restrictions we should imposse on the brane to close the system of Eqs. (\ref{ec1})-(\ref{con1}). However, it is not necessary to 
impose any restriction at all when a given GR perfect fluid solution is considered as a candidate solution for Eqs. (\ref{ec1})-(\ref{con1}). In this case, 
the geometric deformation is minimal and the open system of Eqs. (\ref{ec1})-(\ref{con1}) will be automatically
satisfied, in consequence a BW version of the given GR solution will be automatically generated. The virtue of the MGD approach lies in 
the above fundamental fact, and its usefulness is obvious when physically acceptable GR solutions are investigated in the BW context, as we will see next.
 
Let us start by considering the Tolman IV solution for a perfect fluid in general relativity $(\nu,\lambda,\rho, p)$, which now is {\it deformed}  by five-dimensional effects through $f^{*}(r)$
\begin{equation}\label{tolman00}
e^{\nu}=B^2\,\left(1+\frac{r^2}{A^2}\right),
\end{equation}
\begin{equation}\label{tolman11}
e^{-\lambda}=\frac{\left(1-\frac{r^2}{C^2}\right)\left(1+\frac{r^2}{A^2}\right)}{1+\frac{2\,r^2}{A^2}}+f^{*}(r),
\end{equation}
\begin{equation}\label{tolmandensity}
\rho(r) =\frac{3A^4+A^2\left(3C^2+7r^2\right)+2 r^2 \left(C^2+3 r^2\right)}{8{\pi}C^2\left(A^2+2r^2\right)^2},
\end{equation}
and
\begin{equation}
\label{tolmanpressure} p(r)=\frac{C^2-A^2-3r^2}{8{\pi}C^2\left(A^2+2r^2\right)}.
\end{equation}
In GR, i.e when $f^{*}(r)=0$, $A$, $B$  and $C$ have specific values written in terms of the compactness of the distribution, that is, in terms of $M/R$, 
with $M$ and $R$ the mass and radius of the distribution, which are free parameters satisfying the constraint $M/R<4/9$ [See further bellow Eqs. (\ref{A}-\ref{C})]. However, as it is well known,
in the braneworld scenario the matching conditions are modified, consequently there are five-dimensional effects on these constants which must be considered.
Indeed, in the MGD approach, $A$, $B$  and $C$ in general are functions of the brane tension $\sigma$, being the $\sigma$ dependence determined by matching
conditions. We want to stress that as long as the brane tension $\sigma$ remains constant, $A$, $B$ and $C$ will not be functions of 
the spacetime but functions of the parameters $M$, $R$ and $\sigma$. On the other hand, in general relativity the second fundamental form, which leads to
$p(r)\mid_{r=R}\,=0$ at the stellar surface $r=R$, produces 
\begin{equation}
\label{ABC}
C^2=A^2+3\,R^2.
\end{equation}
We will keep the physical pressure vanishing on the surface, even though this condition may be dropped in the braneworld scenario
\cite{gergely2007}. 
\par
From the point of view of a brane observer, the geometric deformation $f^{*}(r)$ in Eq. (\ref{tolman11}) produced by five-dimensional effects modifies  
the perfect fluid solution [represented by Eqs. (\ref{tolman00})-(\ref{tolmanpressure}) when $f^{*}(r)=0$], introducing thus imperfect fluid effects through the braneworld solution for the geometric function $\lambda(r)$, 
which is obtained using Eqs. (\ref{tolman00}), (\ref{tolmandensity}) and (\ref{tolmanpressure}) in Eq. (\ref{edlrwss}), leading to
\begin{equation}\label{reglambda}
e^{-\lambda(r)}=1-\frac{2\tilde{m}(r)}{r},
\end{equation}
where the interior mass function $\tilde{m}$ is given by
\begin{equation}
\label{massfunction}
\tilde{m}(r)=m(r)-\frac{r}{2}\,f^{*}(r), 
\end{equation}
with $f^{*}(r)$ the {\it minimal geometric deformation} for the Tolman IV solution, given by Eq. (\ref{fsolutionmin}), whose explicit form is obtained using 
Eqs. (\ref{tolman00}), (\ref{tolmandensity}) and (\ref{tolmanpressure}) in Eq. (\ref{fsolutionmin}), hence
%\begin{widetext}
\begin{eqnarray}
\label{gr} f^{*}(r)&=&-\frac{1}{\sigma}\frac{1}{384\pi r(A^{2}+3R^{2})^{2}(2A^{2}+3r^{2})^{3/2}}\left\lbrace (A^{2}+r^{2})\left[ \frac{36\,r\, \sqrt[]{2A^{2}+3r^{2}}}{(A^{2}+2r^{2})^{3}}\left\lbrace 5A^{8}+7A^{6}r^{2}+10A^{2}r^{6} \right. \right. \right. \nonumber \\
&+&12r^{8}+4(6A^{6}+10A^{4}r^{2}-3A^{2}r^{4}-6r^{6})R^{2}+2(15A^{4}+35A^{2}r^{2}+18r^{4})R^{4}\left. \left. \left. \right\rbrace \right. \right. \nonumber \\
&-&216(A^{2}+2R^{2})^{2} \arctan \left( \frac{r}{\sqrt[]{2A^{2}+3r^{2}}} \right)-48\,\sqrt[]{3}(A^{2}+3R^{2})^{2} \log(3r+\sqrt[]{6A^{2}+9r^{2}}) \left. \left. \right] \right\rbrace
\end{eqnarray}
%\end{widetext}
The function $m(r)$ in Eq. (\ref{massfunction}) is the GR mass function,
given by the standard form
\begin{equation}
\label{regularmass2} m(r)=\int_0^r 4\pi
x^2{\rho}dx=\frac{r^{3}(2A^{2}+3R^{2}+r^{2})}{2(A^{2}+3R^{2})(A^{2}+2r^{2})},
\end{equation}
hence the total GR mass is obtained
\begin{equation}
\label{regtotmass} M\equiv
m(r)\mid_{r=R}\,=\frac{R^{3}}{A^{2}+3R^{2}}.
\end{equation}
Finally, the Weyl functions ${\cal P}$ and ${\cal U}$ associated with the geometric deformation shown in Eq. (\ref{gr}), are written as
\begin{equation}
\label{tolmanP}
\frac{\cal P}{\sigma}=\frac{4\,\pi}{3}\frac{(A^4+2 A^2 r^2+2 r^4)}{r^2 (A^2+r^2)^2}\,f^{*}(r)\, ,
\end{equation}
%\begin{widetext}
%\begin{eqnarray}
%\frac{\cal U}{\sigma}&=&\frac{4\pi (A^4+8 A^2 r^2+5 r^4)}{3 r^2 (A^2+r^2)^2}\,f^{*}(r)
%-\frac{(A^4+5 A^2 C^2-3 A^2 r^2+6 C^2 r^2-6 r^4) (3 A^4+3 A^2 C^2+7 A^2 r^2+2 C^2 r^2+6 r^4)}{\sigma\,4 C^4 (A^2+2 r^2)^4} \nonumber \\
%\end{eqnarray}
%\end{widetext}
%\begin{widetext}
\begin{eqnarray}
\label{tolmanU}
\frac{\cal U}{\sigma}&=&\frac{4\pi (A^4+8 A^2 r^2+5 r^4)}{3 r^2 (A^2+r^2)^2}\,f^{*}(r)
\nonumber \\&&
-\frac{1}{\sigma}\frac{9(2 A^4+3 A^2 r^2+2 r^4+3 A^2 R^2+2 r^2 R^2)(2 A^4+A^2 r^2-2 r^4+5 A^2 R^2+6 r^2 R^2)}{4 (A^2+2 r^2)^4 (A^2+3 R^2)^2}\, . \nonumber \\
\end{eqnarray}
%\end{widetext}
The expressions Eq. (\ref{tolman00})-(\ref{tolmanpressure}) along with Eq. (\ref{tolmanP}) and (\ref{tolmanU}) represent an {\it exact analytic solution}
to the system Eq.(\ref{ec1})-(\ref{con1}). We want to emphasize that the expressions for $p$, $\rho$ and $\nu$ in our solution are the same than those 
for the Tolman IV solution, in consequence when these expressions are used in Eq. (\ref{finalsol}), the conditions in Eq. (\ref{H=0}) is obtained.

It can be shown by Eq. (\ref{gr}) that the geometric deformation $f^{*}(r)$ 
depends only on the parameter $A$, which has a well defined expression in terms of the compactness $M/R$ in GR, as will be shown in next section. 
\section{Analysis of the solution}
\subsection{GR case}
In order to see the physicall consequences due to BW, first let us start recalling the GR case, that is, to match the Tolman IV solution to exterior Schwarzschild solution. 
The exterior metric in (\ref{genericext}) will be the Schwarzschild one
\begin{equation}
 e^{\nu^+}=e^{-\lambda^+}=1-\frac{2\,M}{r}\ ,
\end{equation}
therefore at the stellar surface $r=R$ we have
\begin{equation}\label{matchS1}
B^2\,\left(1+\frac{R^2}{A^2}\right)=1-\frac{2\,M}{R}\ ,
\end{equation}
\begin{equation}\label{matchS2}
\frac{\left(1-\frac{R^2}{C^2}\right)\left(1+\frac{R^2}{A^2}\right)}{1+\frac{2\,R^2}{A^2}}=1-\frac{2\,M}{R}\ ,
\end{equation}
whereas the condition (\ref{sfgeneric}) with $f_R^*={\cal U}_R^+={\cal P}_R^+=0$ leads to the expression in Eq. (\ref{ABC}). 
Now by using Eq. (\ref{ABC}) along with Eqs. (\ref{matchS1})-(\ref{matchS2}) the constants $A$, $B$ and $C$ are found in terms of the compactness of the stellar distribution, as shown below
\begin{eqnarray}
\label{A}
A^2/R^2&=&\frac{1-3\,M/R}{M/R}\ ,
\\
\label{B}
B^2&=&1-3\,M/R\ ,
\\
\label{C}
C^2/R^2&=&(M/R)^{-1}\ .
\end{eqnarray}
The values given by Eqs. (\ref{A})-(\ref{C}) guarantee the geometric continuity at $r=R$ when this discontinuity is crossed, for instance,  from the interior geometry, 
described by the Tolam IV solution (\ref{tolman00})-(\ref{tolman11}) ($f^*=0$), to the exterior Schwarszchild geometry, which represents the unique exterior solution for a spherically symmetric distributions in GR. Next we will see that the BW case is quite different.

\subsection{Brane World case}

When a spherically symmetric and static self-gravitating system of radius $r=R$ is considered in the RS BW theory, the space-time $r>R$ surrounding the stellar system, contrary to GR, is not empty but filled by the so-called dark radiation ${\cal U}^+$ 
and dark pressure ${\cal P}^+$. It is well known that, from the point of view of a brane observer, extra dimensional effects modify the Schwarzschild solution through these fields. However, the effects of bulk gravitons on self-gravitating structures 
are not quite understood so far \cite{olps2013}. 

On the other hand, the gravitational collapse of spherically symmetric stellar distributions on the brane could lead to non-static exterior 
\cite{nogo}-\cite{dad}, at least when standard matter is the source of the gravitational field and the configuration has vanishing surface pressure. 
Even in this extreme scenario without Birkhoff theorem, a static exterior is eventually expected. The reason is 
that the (unknown) non-static exterior solution should be transient \cite{nogo}, \cite{ssm}, hence it 
is reasonable to assume that the exterior metric will be static at late times and tend to
Schwarzschild at large distances. However, using more general assumptions than the Oppenheimer-Snyder BW model used in Ref. \cite{nogo}, 
it was found that static exterior can exist for a collapsing star in the radiative bulk scenario \cite{pal}, and also for a near 
dust-like perfect fluid \cite{gergely2007}. Moreover, recently it was proven that a realistic interior solution having a 
very thin dark energy atmosphere can be matched consistently to a 
Schwarzschild exterior \cite{OCG2013}. In summary, the above shows that the presence or not of static black holes remains an open issue in BW.
%showing thus that the searching of extradimensional effects by observational data could be more difficult than expected.

Since the exterior spacetime of a spherically symmetric configurations remains unknown on the brane due to the lack of a 5-dimensional solution, there are many ways to modify the Schwarzschild solution, i.e., 
there are many black hole solutions for a spherically symmetric static ``vacuum'' in 5-dimensional 
gravity in the RS BW scenario \cite{FW11}-\cite{LBH13}, \cite{dadhich}, \cite{CFMsolution}. Next, regarding the weak-field limit in the BW, an approximate exterior solution is developed and considered in the matching conditions at the stellar surface $r=R$.

\subsection {Far-field correction to the Newtonian potencial}

As we have mentioned, the Weyl stresses imply that the exterior solution for a 
spherically symmetric distribution is no longer the Schwarzschild metric and therefore there are many possible solutios to the effective four-dimensional
vacuum Einstein equations~\cite{SMS}, namely any metric such that
\be
\label{generalvacuum}
R_{\mu\nu}-\frac{1}{2}\,g_{\mu\nu}\,R^\alpha_{\ \alpha}
=
\mathcal{E}_{\mu\nu}
\qquad
\Rightarrow
\qquad
R^\alpha_{\ \alpha}
=0
\ .
\ee

The solution to Eq. (\ref{generalvacuum}) must satisfy the weak-field limit \cite{tanaka}, which is given by
\begin{equation}
\label{newton}
\Phi \sim\, -\frac{G\,{\cal M}}{r}\left(1+\frac{2\, \ell^2}{3\,r^2}\right)\ ,
\end{equation}
where $\ell$ is the curvature radius of $AdS_5$. Unfortunately, none of 
the few known analytical solutions to Eq. (\ref{generalvacuum}) satisfy the weak-field limit in Eq. (\ref{newton}) and therefore cannot describe the end-state of collapse. Indeed, to our knowledge, an exact spherically symmetric 
exterior solution on the brane satisfying the weak-field limit in Eq. (\ref{newton}) remains unknown so far. For instance, while it is true that the well known tidaly charged metric found in Ref. \cite{dadhich} shows the correct 5D behaviour of the potential at short distances, and therefore could be a good 
approximation in the strong-field regime for micro black holes \cite{covalle1}, \cite{covalle2}, the astrophysical scenario is quite different. Likewise, 
although the vacuum braneworld solution found by Casadio, Fabbri
and Mazzacurati in Ref. \cite{CFMsolution} is tremendously useful to elucidate the specific role played for 
both Weyl functions \cite{olps2013}, it does not satisfy the limit in Eq. (\ref{newton}). Moreover, its condition of null dark energy 
is too strong and therefore this solution should be condidered just an useful (unphysical) toy model in the astrophysical scenario. (For a resent study regarding this solution in the bulk, see Ref. \cite{roldaocqg2013}).

Since we want to elucidate the effects of bulk gravitons on stellar structure, and we lack an exterior solution, the potential in  Eq. (\ref{newton}) could be  
helpful to obtain some relevant information. For instance, it is reasonable to assume that the unknow 4-dimensional solution should be 
close to the solution associated to the potential in Eq. (\ref{newton}) ($G=1$)
\begin{equation}
 \label{aprox00}
g^{+}_{00} \sim\, 1-2\,{\cal M}\left(\frac{1}{r}+\frac{2 \ell^2}{3\,r^3}\right)\ ,
\end{equation}
\begin{equation}
 \label{aprox11}
\left(g^{+}_{11}\right)^{-1} \sim\, 1-2\,{\cal M}\left(\frac{1}{r}+\frac{ \ell^2}{3\,r^3}\right)\ .
\end{equation}
In this approximation, the exterior Weyl fluid behaves
%\begin{equation}
%\frac{6\,{\cal P}^+}{k^2\sigma} \sim (l^2 M (84 M^2-91 M r+25 r^2))/(3 r^5 (-2 M+r)^2)
%\end{equation}
%\begin{equation}
%\frac{6\,{\cal U}^+}{k^2\sigma} \sim\, -((2 l^2 M (36 M^2-37 M r+10 r^2))/(3 r^5 (-2 M+r)^2))
%\end{equation}
\begin{equation}
\label{approxpp}
\frac{6\,{\cal P}^+}{k^2\sigma} \sim \frac{{\cal M}\ell^2}{3r^5}\left[\frac{84({\cal M}/r)^2-91({\cal M}/r)+25}{(1-2{\cal M}/r)^2}\right]\ ,
\end{equation}
\begin{equation}
\label{approaxuu}
\frac{6\,{\cal U}^+}{k^2\sigma} \sim -\frac{2{\cal M}\ell^2}{3r^5}\left[\frac{36({\cal M}/r)^2-37({\cal M}/r)+10}{(1-2{\cal M}/r)^2}\right]\ ,
\end{equation}
where the expressions in Eqs. (\ref{aprox00})-(\ref{aprox11}) have been used in Eq. (\ref{pp})-(\ref{uu}) [with $p = \rho = 0$].

Therefore when the deformed Tolman IV interior solution, given by Eqs. (\ref{tolman00}) and (\ref{reglambda}), is used 
along with the exterior solution in Eqs. (\ref{aprox00})-(\ref{aprox11}) in the matching conditions (\ref{ffgeneric1}) and (\ref{ffgeneric2}), we have
\begin{equation}
\label{aproxmatch00}
B^2\,\left(1+\frac{R^2}{A^2}\right)\sim\,1-\frac{2\cal{M}}{R}-\frac{4\,{\cal M}\ell^2}{3\,R^3}\ ,
\end{equation}
\begin{equation}\label{aproxmatch11}
{\cal M} \sim \frac{M-\frac{R}{2}f^{*}_R}{1+\frac{\ell^2}{3\,R^2}} \sim {M} -\frac{R}{2}\,f^{*}_R-\frac{M\ell^2}{3R^2}\ , 
\end{equation}
where in Eq. (\ref{aproxmatch11}) the approximation $f^{*}_R\, (\ell/R)^2\sim\,\sigma^{-1}(\ell/R)^2\sim\,0$ has been used. 

Then when Eq. (\ref{aproxmatch11}) is used in the condition (\ref{aproxmatch00}), we obtain
\begin{equation}
\label{sup1}
B^2\,\left(1+\frac{R^2}{A^2}\right) \sim 1-\frac{2\,{M}}{R}+{\bar f}^*_R\ ,
\end{equation}
where the bar-function
\begin{equation}
\label{fbarr}
{\bar f}^*_R \equiv\, f^*_R-\frac{2\,{M}\ell^2}{3\,R^3}
\end{equation}
represents the bulk effects on the righ-hand side of Eq. (\ref{matchS1}). These effects can be written in terms of the brane tension $\sigma$ or the curvature radius of the bulk $\ell$ when the second fundamental form (\ref{sfgeneric}) is used, that is, by using Eqs. (\ref{approxpp})-(\ref{approaxuu}) and $p_R = 0$ in the condition (\ref{sfgeneric}), thus a relationship between the geometric deformation $f^{*}(r)$ and the curvature radius of the bulk $\ell$ is found at the stellar surface $r=R$ as
\begin{equation}
 \label{sfgenericapproxim}
{f^*_R} \sim\frac{10 {\cal M} \ell^2}{3 R^3 }\left(\frac{1-2M/R}{1-2{\cal M}/R}\right)
\left(1-\frac{8{\cal M}}{5R}\right)+{\cal O}(\ell^4/R^4)
\sim\frac{10 {M} \ell^2}{3 R^3 }\left(1-\frac{8{M}}{5R}\right)\ ,
\end{equation}
in consequence the expression in Eq. (\ref{fbarr}) may be written as
\begin{equation}
\label{fbarr2}
{\bar f}^*_R  \sim \frac{8}{3}\left(1-\frac{2\,M}{R}\right)\frac{M \ell^2}{R^3} 
\sim \frac{4}{5}\frac{\left(1-\frac{2M}{R}\right)}{\left(1-\frac{8M}{5R}\right)}f^{*}_R\ ,
\end{equation}
showing thus that the bar-function in Eq. (\ref{fbarr2}) is allways positive.

The expression (\ref{sup1}) clearly shows that the values of $A$ and $B$ cannot 
be that from Eqs. (\ref{A}) and (\ref{B}) because that values would lead to the trivial Schwarzschild condition ${\bar f}^*_R=0$. 
Therefore the GR values of $A$ and $B$, hereafter called $A_{0}$ and $B_{0}$, have been modified by five-dimensional effects and 
cannot be constants anymore but functions of the brane tension $\sigma$ [or the curvature radius of the bulk $\ell$, according to Eq. (\ref{fbarr2})], 
as will be shown next. 

The expression in (\ref{sup1}) represents 
a condition which must be used to find two unknown functions $A$ and $B$, hence the problem at the surface seems not closed and 
therefore additional information should be added. However, we will see that no additional information is needed, as explained below.

First of all, the BW effects on the GR condition (\ref{matchS1}) is explicitly shown in the right-hand side of Eq. (\ref{sup1}), showing 
that these modifications are proportionals to the geometric deformation $f^*_R$, which is proportional to $\sigma^{-1}$ [see Eq. (\ref{gr})]. Therefore the unknown functions $A$ and $B$ can be written as
\begin{eqnarray}
\label{dA}
 A = A_{0}+\delta\,A\ ,
\\
\label{dB}
 B = B_{0}+\delta\,B\ ,
\end{eqnarray}
where $\delta$ represents the modification due to five-dimensional effects, which are functions of the brane tension $\sigma$. 
Hence, the problem is reduced to 
finding the unknown $\delta$ functions in Eq. (\ref{dA})-(\ref{dB}) by using the condition in Eq. (\ref{sup1}). On the other hand, since the 
constants shown in Eqs. (\ref{A})-(\ref{C}) are modified by bulk gravity effects, this must occure by a change in the compactness $M/R$, as clearly shows the 
right-hand side of Eqs. (\ref{A})-(\ref{C}), but $R$ is a constant free parameter, therefore the five-dimensional effects on $A$ and $B$ are produced by 
bulk gravity effect $\delta\,M$ on GR mass $M_0$
\begin{eqnarray}
\label{dM}
 M = M_{0}+\delta\,M\ ,
\end{eqnarray}
hence $\delta\,A$ and $\delta\,B$ have the {\it same source} and are not independent. Therefore all we need to do is to find $A=A(B)$ by using the compactness as a common variable, as shown 
by Eqs. (\ref{A}) and (\ref{B}), hence
\begin{equation}
 \label{AB}
 A^2=3\,R^2\frac{B^2}{(1-B^2)}\ ,
\end{equation}
in consequence the problem 
at the surface is closed. Clarified this point, next step is to examine the five-dimensional effects on the physical variables. 
For instance, to see bulk gravity consequences on the pressure $p$, all 
we have to do is to use Eq. (\ref{dA}) in Eq. (\ref{tolmanpressure}), hence the modification $\delta\,p$ will be found. Therefore we need to 
determined $\delta\,A$ in Eq. (\ref{dA}), as shown below.

Using Eqs. (\ref{dA})-(\ref{dB}) in (\ref{sup1}), yields  
\begin{equation}
\label{sup2}
(B_{0}+\delta\,B)^2\,\left[1+\frac{R^2}{(A_{0}+\delta\,A)^2}\right] \sim\, 1-\frac{2\,{M_0}}{R}+{\bar f}^*_R\ ,
\end{equation}
where $M_0$ is used to stress that the $M$ in Eq. (\ref{sup1}) actually is the GR value of $M$. Now keeping in 
Eq. (\ref{sup2}) linear terms in $\delta$, we have
\begin{equation}\label{sup3}
B_0^2\,\left(1+\frac{R^2}{A_0^2}\right)+2\,B_0\left(1+\frac{R^2}{A_0^2}\right)\delta\,B-2\frac{\,B_0^2\,R^2}{A_0^3}\delta\,A \sim\, 1-\frac{2\,M_0}{R}+{\bar f}^*_R\ ,
\end{equation}
and by using Eq (\ref{matchS1}) [GR case, where $B=B_0$] in Eq. (\ref{sup3}) we obtain
\begin{equation}\label{sup4}
2\,B_0\left(1+\frac{R^2}{A_0^2}\right)\delta\,B-2\frac{\,B_0^2\,R^2}{A_0^3}\delta\,A \sim\, {\bar f}^*_R\ .
\end{equation}
In order to find $\delta\,A$ in Eq. (\ref{sup4}) $\delta\,B$ must be determined. To accomplish this the expression in Eq. (\ref{AB}) is used, yielding
\begin{equation}
 \label{dAB}
\delta\,A=\frac{3\,R^2}{A}\frac{B}{(1-B^2)^2}\,\delta\,B \ .
\end{equation}
Using Eq (\ref{dAB}) in Eq. (\ref{sup4}) leads to
\begin{equation}
 \label{sup5}
\delta\, A(\sigma)\sim\,\frac{A_0^3}{4\,R^2\,B_0^4}\,{\bar f}^*_R\ ,
\end{equation}
therefore the function $A$ in Eq. (\ref{dA}) is written as
\begin{equation}
 A(\sigma)\sim\,A_0+\frac{A_0^3}{4\,R^2\,B_0^4}\,{\bar f}^*_R(\sigma)+{\cal O}(\sigma^{-2})\ .
\end{equation}

At this stage, we have all the necessary tools needed to examine the five-dimensional effects on the physical variables. 
For instance, to see bulk gravity consequences on the pressure $p(r,\sigma)$, we rewrite Eq. (\ref{tolmanpressure}) as
\begin{equation}
\label{tolmanpressure2} 
p(r,\sigma)=\frac{3(R^2-r^2)}{8{\pi}\left(A^2+3\,R^2\right)\left(A^2+2r^2\right)}\ .
\end{equation}
As the bar-function ${\bar f}^*_{R}$ in Eq. (\ref{fbarr2}) is positive, then from Eq. (\ref{sup5}) we can see that $\delta\,A>0$, 
in consequence is straightforward to see that the pressure in Eq. (\ref{tolmanpressure2}) is always reduced by five-dimensional effects. 

Finally, by using Eqs. (\ref{A})-(\ref{B}) and Eq. (\ref{dAB}) in Eq. (\ref{sup5}) $\delta\,M$ may be written as 
\begin{equation}
\label{deltaM}
\delta\,M(\sigma) \sim\, -\frac{R}{2}\, {\bar f}^*_R\ , 
\end{equation}
hence the bulk effects on the compactness $M/R$ may be expressed as
\begin{equation}
\label{deltaMR}
\delta\,[M(\sigma)/R]\sim\,-{\bar f}^*_R/2\ , 
\end{equation}
or, according to Eq. (\ref{fbarr2}), in terms of the brane tension $\sigma$ or curvature radius of the bulk $\ell$
\begin{equation}
\label{jl}
\delta\,[M(\sigma)/R]  \sim -\frac{4}{3}\left(1-\frac{2\,M}{R}\right)\frac{M \ell^2}{R^3} 
\sim -\frac{2}{5}\frac{\left(1-\frac{2M}{R}\right)}{\left(1-\frac{8M}{5R}\right)}f^{*}_R\ .
\end{equation}

The pressure in (\ref{tolmanpressure2}) may be written in terms of the compactness as
\begin{equation}
\label{tolmanpressure3} 
p(r,\sigma)=\frac{3(1-r^2/R^2)}{8{\pi}\,R^2\left[1-\left(3-2r^2/R^2\right)\frac{M(\sigma)}{R}\right]}\left[\frac{M(\sigma)}{R}\right]^2\ ,
\end{equation}
where $M(\sigma)$ in Eq. (\ref{tolmanpressure3}) is given by
\begin{equation}
 M(\sigma) = M_0 + \delta\,M(\sigma)\ .
\end{equation}

Expressions in Eq. (\ref{deltaMR}) and (\ref{tolmanpressure3}) clearly shows that BW consequences on pressure occure through bulk gravitons effects on the compactness of the stellar structure. The result shown by Eq. (\ref{jl}) strongly suggests that the effects of bulk gravitons on stellar configurations 
act in such a way that always reduce the compactness, in agreement with the results found in Ref. \cite{germ}.

\section{Conclusions}

In this paper, an exact analytic interior solution to four-dimensional effective Einstein's field equations
for a  non uniform stellar structure was found in the context of the Randall-Sundrum braneworld. 
By using this analytic solution, an exhaustive analysis of the extra-dimensional consequences on realistic stellar interiors is developed, 
finding strong evidences in favor of the hypothesis that compactness and pressure are always reduced due to bulk effects on stellar configurations. 

The interior solution was constructed from a well known 
spherically symmetric stellar solution for a perfect fluid in GR, namely, the Tolman IV solution. 
In order to produce the braneworld version containing the anisotropic effects necessary for realistic stellar models, 
the MGD approach was used to modify the perfect fluid solution represented by the Tolman IV solution, 
thus generating exact analytic expressions for the Weyl fields on the brane, namely, the scalar ${\cal U}$ and anisotropy ${\cal P}$. 

As the Tolman IV solution is a solution to $4D$ Einstein's field equations in GR, it removes all the 
non-local sources from the geometric deformation $f(\nu,\rho,p)$ in the generic expression given by Eq. (\ref{edlrwss}), leaving only the high energy terms shown explicitly in Eq. (\ref{fsolutionmin}), which are quadratic terms in the density and pressure. Hence the higher the density, the more geometric deformation will be produced, and as a consequence the anisotropy induced will be higher for more compact 
distributions, as can easily seen through Eq. (\ref{tolmanP}). Finally, we want to stress that, while it is true that both the pressure and density in (\ref{tolmandensity}) and (\ref{tolmanpressure}) are modified through the change $A\rightarrow\;A(\sigma)$, their physical acceptability is not lost, given that is inherited from the Tolman IV solution. In other words, the deformation undergone by the density and pressure is not enough to jeopardize the physical acceptability of the BW system.
 
On the other hand, since we lack an exterior solution, 
the far-field correction to the Newtonian potencial in the BW was used to construct an exterior geometry associated with this potential. 
In this approximation, it was found that bulk effects always reduce the compactness of stellar configurations, in agreement with the conjectured hypothesis in Ref. \cite{germ}.

The analytic four-dimensional solution developed in this paper represents the point of view of a
brane observer, hence it is not known whether the bulk eventually constructed will be free of singularities. 
Despite the above, it was found that the MGD principle represents a powerful tool in the searching of analytic 
solutions in the braneworld context. Hence it could be useful in 
the study of the five-dimensional geometry. Indeed, we could use Campbell-Magaard theorems \cite{campbell, sss} 
to extend the generic solution represented by the MGD metric in (\ref{mgdmetric}) through the bulk, locally at least. 
Also it could be investigated the consequences of the MGD metric in (\ref{mgdmetric}) on five-dimensional 
bulk by introducing an extra dimension $y$ dependence in the MGD metric, similar to the study developed in Ref. \cite{kanti2002}. All this certainly deserves further investigation \cite{kan}.

%------------------------------------------------------------------------------------------------

%-------------------------------------------------------------------------------------------------

\begin{thebibliography}{99}
\bibitem{phymeaning} That is, regular at the origin, pressure and density
defined positive, well defined mass and radius, monotonic decrease
of the density and pressure with increasing radius, dominant energy condition satisfied, subliminal sound speed, etc.
\bibitem{kan} Once a extra dimension $y$ dependence in the MGD metric is introduced, it could be find the kind of bulk matter giving rise to such $y$-dependent solution.
\end{thebibliography}
\end{document}